\documentclass[journal]{IEEEtran}

\ifCLASSINFOpdf

\else

\fi

\usepackage{amsmath}
\usepackage{graphicx}
\usepackage{amsfonts,amssymb}
\usepackage{mathrsfs}
\usepackage{tabularx}
\usepackage{enumerate}
\usepackage{booktabs}
\usepackage{algorithm}
\usepackage{algorithmic}
\usepackage{caption}
\usepackage{subfigure}
\usepackage{amsfonts}
\usepackage{cite}
\usepackage{tabularx}
\usepackage{amsthm}
\usepackage{booktabs}
\usepackage{mathrsfs}
\usepackage{mathrsfs}
\usepackage{caption}
\usepackage{amsmath}

\newtheorem{lemma}{Lemma}
\newtheorem{myDef}{Definition}
\newtheorem{myTheo}{Theorem}
\begin{document}

\title{Placement Delivery Array Design via Attention-Based Deep Neural Network}


\author{\IEEEauthorblockN{Zhengming Zhang, Meng Hua, Chunguo Li, ~\IEEEmembership{Senior Member,~IEEE},\\ Yongming Huang, ~\IEEEmembership{Senior Member,~IEEE}, Luxi Yang, ~\IEEEmembership{Member,~IEEE}}}
\maketitle

\begin{abstract}
A decentralized coded caching scheme has been proposed by Maddah-Ali and Niesen, and has been shown to alleviate the load of networks. Recently, placement delivery array (PDA) was proposed to characterize the coded caching scheme. In this paper, a neural architecture is first proposed to learn the construction of PDAs. Our model solves the problem of variable size PDAs using mechanism of neural attention and reinforcement learning. It differs from the previous attempts in that, instead of using combined optimization algorithms to get PDAs, it uses sequence-to-sequence model to learn construct PDAs. Numerical results are given to demonstrate that the proposed method can effectively implement coded caching. We also show that the complexity of our method to construct PDAs is low.
\end{abstract}


\IEEEpeerreviewmaketitle
\section{Introduction}

\renewcommand{\thefootnote}{}
\footnotetext{Z. Zhang, M. Hua, C. Li, Y. Huang, and L. Yang are with the National Mobile Communications Research Laboratory, School of Information Science and Engineering, Southeast University, NanJing 210096, P. R. China (email: zmzhang, mhua, chunguoli, huangym and lxyang@seu.edu.cn).}

\IEEEPARstart{D}{ue} to the exponential growth in the number of smart mobile equipments and innovative high-rate mobile data services (such as videos streaming for mobile gaming and road condition monitoring), 5G networks should accommodate the overwhelming wireless traffic demands \cite{7010538}, \cite{7456186}. Deploying intelligent caching is efficient and able to cope with the demands of users for it can quickly obtain the information they required \cite{7914647}, \cite{7883826}.

The gain from traditional (uncoded) caching approaches derives from making content available locally, and it is constrained by the limited memory available at each individual user. In the seminal work \cite{6763007}, a coded caching scheme was proposed for the centralized caching system by Maddah-Ali and Niesen which is referred to AN scheme in this paper. The AN scheme can create multicast opportunities depending on the cumulative memory available at all users. It has been used in many scenarios, for example, \cite{7842115}  proposed two schemes in
finite file size regime, and \cite{8017548} studied coded caching method for wireless networks with unequal link rates. However, in order to implement the coded caching scheme proposed in \cite{6763007}, each file must be split into $F$ file packages. The number of packages generally increases exponentially with the number of users. In order to reduce $F$, placement delivery array (PDA) was proposed creatively to describe placement and delivery phase \cite{7973185}.

It¡¯s completely equivalent to construct a PDA and design a coded caching scheme, because a PDA can indicate what should be cached by users and what should be sent by the server in a single array. Numerous methods have been proposed to address the construction of PDAs. The method proposed in \cite{7973185} significantly decreases $F$, while only suffering from a slight sacrifice. The PDA model was reconstructed from the perspective of graph theory in \cite{DBLP:journals/corr/ShangguanZG16}, and a hypergraph theoretical approach was proposed to establish connection between graph theory and coded caching. The authors in \cite{8080217} found the connection between strong edge coloring of bipartite graphs and PDA construction and they proposed a placement delivery array design algorithm using graph theory presented in \cite{Quinn2015Strong}.

However, the problem of finding an optimal edge coloring is NP-hard and the fastest known algorithms for it take exponential time. Although new PDAs can be discovered through finding strong edge coloring, for the general case, only some sporadic results exist. In this paper, we revisit PDA design problem in a simpler perspective, \emph{i.e.}, sequence-to-sequence (Seq2Seq) learning model, which is widely used in natural language processing \cite{Sutskever2014Sequence}, \cite{8246560}. This inspires us to present a deep neural architecture to devise coded caching schemes. The architecture has three key technologies, \emph{i.e.}, Seq2Seq learning \cite{Sutskever2014Sequence}, content based input attention \cite{Bahdanau2014Neural} and reinforcement learning.

The main contributions of this paper are summarized as follows:

(i) A deep neural architecture is first proposed to learn the construction of PDAs, and it allows us to realize coded caching using deep learning technology.

(ii) Attention model is used to deal with the fundamental problem of representing variable size of PDAs. Reinforcement signals are used to accelerate deep neural network training.
Our results demonstrate that this approach can achieve approximate solutions to the problems of construction of PDAs that are computationally intractable.

The rest of this paper is organized as follows. In Section II, we introduce the system model and the
backgrounds of PDA. In Section III we propose the attention-based Seq2Seq learning algorithm of the construction of PDAs. Finally, numerical results and discussion are presented in Section IV, and a conclusion is reached in Section V.

\section{SYSTEM MODEL AND BACKGROUND}
We consider a caching system composing of one server and $N$ files ${{\cal W}} = \{ W_1 ,W_2 , \cdots ,W_N \}$. This server is connected to $K$ users through an error-free shared link, and the set of all users is denoted by \({\cal K} = \{ 1,2, \cdots ,K\}\) $(N > K)$. We assume that each file has equal size, and each user is equipped with a cache of size $M$. The caching system is parameterized by $K$, $M$ and $N$, and it is called a $(K, M, N)$ caching system. According to the AN scheme, the caching system has two phases:

\textbf{Placement Phase: }A file is subdivided into $F$ equal packets, \emph{i.e.}, $W_i=\{W_{i,j}:j\in[1,F]\}$. The size of each packet is $1/F$. These packets are placed in users¡¯ cache memories deterministically independent of the user demands which are assumed to be arbitrary. This phase is performed during off-peak times.

\textbf{Delivery Phase: }Each user randomly requests one file from $\cal W$ independently. Their requests constitute $\textbf{\emph{d}}=(d_1,d_2,\cdots,d_K)$, where $d_k$ means that user $k\in \cal K$ requests the file $W_{d_k}$ for any $d_k\in [1,N]$. Once the server received $\textbf{\emph{d}}$ it broadcasts a coded signal of at most $RF$ (where $R$ is called the delivery rate) packets to users, such that each users can recover its requested file from the signal received with the help of the contents received in the placement phase.

The goal is to minimize the load of $RF$ packets. The AN model can be reformulated as a PDA design problem.

\begin{myDef}[Placement Delivery Array, \cite{7973185}]
For positive integers $K$, $F$ and nonnegative integers $Z$ and $S$ with $F\geq Z$, an $F\times K$ array $\textbf{\emph{P}} = (p_{i,j})$, $i\in[1,F]$, $j\in[1,K]$, composed of a specific symbol $\ast$ and $S$ nonnegative integers $1,2,\cdots,S$, is called a $(K,F,Z,S)$ placement delivery array (PDA) if it satisfies the following conditions:

C1. The symbol $\ast$ appears $Z$ times in each column;

C2. For any two distinct entries $p_{i_1,j_1}$ and $p_{i_2,j_2}$ , $p_{i_1,j_1} = p_{i_2,j_2} = s$ is an integer only if

\qquad a. $i_1\neq i_2$, $j_1\neq j_2$, \emph{i.e.}, they lie in distinct rows and distinct columns; and

\qquad b.  $p_{i_1,j_2} = p_{i_2,j_1} = \ast$, \emph{i.e.}, the corresponding $2\times2$ subarray formed by rows $i_1$, $i_2$ and columns $j_1$, $j_2$ must be of the following form
\[
\left( {\begin{array}{*{20}c}
   s & *  \\
   * & s  \\
\end{array}} \right){\rm{  or  }}\left( {\begin{array}{*{20}c}
   * & s  \\
   s & *  \\
\end{array}} \right)
\]

\end{myDef}

Based on a $(K,F,Z,S)$ PDA $\textbf{P}$, caching system with $M/N = Z/F$ can be obtained as follows:

1. \textbf{Placement Phase: } Each file is split into $F$ packets,\emph{i.e.}, $W_i={W_{i,j}:j\in[1,F], \forall i \in [1,N]}$, and user $k$ caches packets
\begin{equation}
C_k  = \{W_{i,j}: p_{j,k}=\ast, \forall i \in [1,N]\}.
\end{equation}

2. \textbf{ Delivery Phase: } The server receives the request $\textbf{\emph{d}}$, at the time slot $s$, it broadcasts:
\begin{equation}
\mathop  \oplus \limits_{p_{j,k}  = s,j \in [1,F],k \in [1,K]} W_{d_k ,j},
\end{equation}
where the operation $\oplus$ is bitwise Exclusive OR (XOR) operation.

The authors of \cite{8080217} review the definitions from graph theory and they found the connections between strong edge coloring of bipartite graphs and PDAs.

\begin{myDef}
For a bipartite graph $G(\mathscr{K},\mathscr{F},\mathscr{E})$, where$\mathscr{K}$ and $\mathscr{F}$  are the disjoint vertex sets, $\mathscr{E}$ is the edge set. The degree of a vertex is the number of edges incident to the vertex, with loops counted twice; a strong edge coloring of $G$ is an assignment of colors to edges such that, any pair of edges with same color are neither adjacent to each other nor adjacent to any third edge in $\mathscr{E}$.
\end{myDef}

Then the relation between PDA $\textbf{P}$ and  bipartite graph $G(\mathscr{K},\mathscr{F},\mathscr{E})$ is given by Lemma 1.

\begin{lemma}
The array $\textbf{\emph{P}}$ composed of symbol $\ast$ and $1,2,\cdots,S$ is a PDA if and only if its corresponding colored bipartite graph $G(\mathscr{K},\mathscr{F},\mathscr{E})$ satisfies

1. The vertices in $\mathscr{K}$ has a constant degree;

2.  The corresponding coloring is a strong edge coloring.
\end{lemma}

\emph{Proof :} Please refer to \cite{8080217}.

Lemma 1 shows that PDA construction suffers from a high complexity. Our work is to reduce the complexity using Seq2Seq model and the following conclusion:
\begin{myTheo}
Assume a PDA $\textbf{\emph{P}}$ corresponds to the colored bipartite graph $G(\mathscr{K},\mathscr{F},\mathscr{E})$, and  each vertex in $\mathscr{K}$ has the constant degree $\Delta=F-Z$. For each vertex $v\in \mathscr{K}$, randomly select $\delta (0<\delta<\Delta)$ edges, then we get a new colored bipartite graph $\widetilde{G}$ and its corresponding array $\widetilde{\textbf{\emph{P}}}$ is a PDA.
\end{myTheo}

\emph{Proof :} From Lemma 1, we know that the corresponding coloring of $G(\mathscr{K},\mathscr{F},\mathscr{E})$ is a strong edge coloring, thus any adjacent edges in $\widetilde{G}$ have different colors, and any edges adjacent to any third edge also have different colors. The vertices in $\widetilde{G}$ has a  constant degree $\delta$, and the bipartite graph $\widetilde{G}$ is strong edge colored graph, thus $\widetilde{\textbf{P}}$ is a PDA.

Obviously, Theorem 1 does not guarantee that the resulting PDA is optimal, but it is useful for our learning model because it can expand our trainable data set.

\textbf{Seq2Seq learning problem for PDA design :} From AN scheme and Lemma 1, we can find that the placement phase basing on a $(K,F,Z,S)$ caching system whose PDA is \textbf{P} can produce a $F\times K$ adjacency matrix $A=(a_{i,j})$, where $a_{i,j}=1$, if $p_{i,j}\neq \ast$ and $a_{i,j}=\rm{Inf}$, if $p_{i,j}= \ast$. Define $E=(e_1,e_2,\cdots,e_L)$ as an ordered sequence compose of adjacent edges in $A$, where $e_l=(i,j), a_{i,j}=1$ and $L$ is the number of edges of $A$. Assume we have another sequence $C=(c_1,c_2,\cdots,c_L)$, where $c_l\in \{1,2,\cdots,S\}$ is the color of the edge $e_l$. Then we can generate an array $\overline {\textbf{P}}=(\overline {p}_{i,j})$
\begin{equation}
\overline {p}_{i,j}  = \left\{ \begin{array}{l}
 c_l ,\; if \; a_{i,j}=1 \\
 *,\; if \; a_{i,j}=\rm{Inf}. \\
 \end{array} \right.
\end{equation}
The Seq2Seq learning problem for PDA design is that given sequence $E$ we should find sequence $C$ so that the array $\overline {\textbf{P}}$ is a PDA.

\emph{Remark 1:} Different $(K,F,Z,S)$ caching system has different size of the PDA, thus the size of output dictionary of the sequence $C$ depends on the length of the input sequence $E$. Traditional Seq2Seq learning methods require the size of the output dictionary to be fixed. Therefore, we cannot directly apply this framework to the PDA design problem.

\section{SOLVE PDA LEARNING PROBLEM}
We first review the Seq2Seq and input-attention models that are the baselines for this work, and then describe our model using attention like \cite{Vinyals2015Pointer} and reinforcement learning like \cite{Yu2017Seqgan}.

\begin{figure*}
\centering
\resizebox{12.0cm}{4.2cm}{\includegraphics{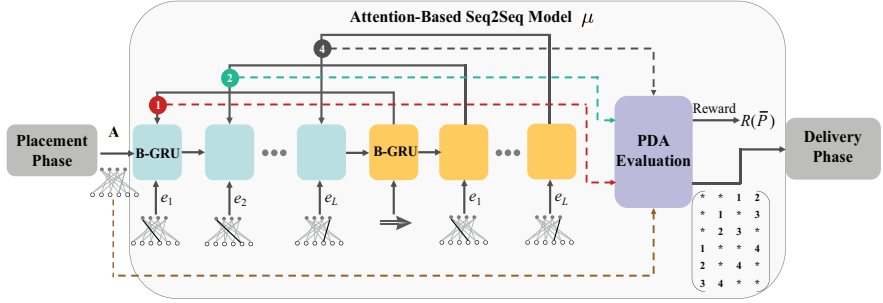}}
\caption{Attention-Based Seq2Seq Placement Delivery Network}
\end{figure*}

\textbf{Seq2Seq model: }Assume we have a training sequence pair, $(E,C^E)$, the Seq2Seq model computes the conditional probability $p(C^E|E;\theta)$. A learnable model with parameters $\theta$  (in this paper we use gated recurrent unit, \emph{i.e.}, GRU) is used to estimate the terms of the probability chain rule
\begin{equation}
p(C^E |E;\theta ) = \prod\limits_{i = 1}^L {p_\theta  (C_i |C_1 , \ldots ,C_{i - 1} ,E;\theta )},
\end{equation}
where $E=(e_1,e_2,\cdots,e_L)$ is a sequence of $L$ vectors and $C^E=(c_1,c_2,\cdots,c_L)$  is a sequence of $L$ indices, each $c_l$ belongs to $\{1,2,\cdots,S\}$. If we take \({\rm K}\) samples from training set, we maximize the conditional probabilities to learn the parameters of the model, \emph{i.e.}
\begin{equation}
\theta^{*} = \mathop {\arg \max }\limits_\theta  \sum\limits_{k = 1}^{\rm K} {\log p(C_k^E |E_k ;\theta )}.
\end{equation}
We use a GRU to model ${p_\theta  (C_i |C_1 , \ldots ,C_{i - 1} ,E;\theta )}$. A GRU is formulated as
\begin{equation}
\left\{ \begin{array}{l}
 r_t  = \sigma (U^r x_t  + W^r y_{t - 1}  + b^r ), \\
 z_t  = \sigma (U^z x_t  + W^z y_{t - 1}  + b^z ), \\
 \widetilde{y}_{t}  = g(U^s x_t  + r_t  \circ W^s y_{t - 1}  + b^z ), \\
 y_t  = z_t  \circ y_{t - 1}  + (1 - z_t )\circ \widetilde{y}_{t},  \\
 \end{array} \right.
\end{equation}
where $x_t$ is the input variable at time $t$, $U^{*}$ and $W^{*}$ are the weight matrices applied on input and hidden units, respectively; $\sigma(\cdot)$ and $g(\cdot)$ are sigmoid and tangent activation functions, respectively; $b^{*}$ is the bias, $y_t$ is the output and $\circ$ means element-wise product. The GRU is fed $E_t$ at each time step $t$ until the end of the input sequence is reached, at which time a special symbol $\Rightarrow$ is input to the model. The model then switches to the generation mode. This model has computational complexity of ${\cal{O}}(n)$ under the assumption that the number of outputs is ${\cal{O}}(n)$.

\textbf{Attention model: }Seq2Seq model constrains the amount of information and computation that can arrive at any part of the generative model. This problem can be ameliorated by using attention model. The attention vector at time step $t$ is given by
\begin{equation}
\left\{ \begin{array}{l}
 u_j^t  = \beta ^T \tanh (W^1 s_j  + W^2 d_t ), \\
 a_j^t  = \rm{softmax} (u_j^t ), \\
 d_t^{'}  = \sum\limits_{j = 1}^L {a_j^t s_j },  \\
 \end{array} \right.
\end{equation}
where $j\in\{1,2,\cdots,L\}$; $(s_1,\cdots,s_L)$ and $(d_1,\cdots,d_L)$ are the encoder and decoder hidden states, respectively; $u_t$ is the attention mask over the inputs; $\beta$, $W^1$ and $W^2$ are learnable parameters. Usually, $d_t^{'}$ and $d_t$ are used as hidden states which are fed to the next time step in the attention model.

\textbf{Our model: }We would like the color of each edge to process not only the preceding edges, but also the following edges. Hence, we use a bidirectional GRU (B-GRU) which consists of forward and backward GRUs. The forward GRU reads the input sequence
as it is ordered and calculates the forward hidden states $(\overrightarrow{s_1},\cdots,\overrightarrow{s_L})$. The backward GRU reads the sequence in the reverse order and calculates the backward hidden states $(\overleftarrow{s_1},\cdots,\overleftarrow{s_L})$. Finally, we obtain the hidden states $s$ by  concatenating the forward hidden state and the
backward one, \emph{i.e.}, $s_l=[\overrightarrow{s_l};\overleftarrow{s_l}]$.

As mentioned in Remark 1, for the PDA design problem ,the output size is related to the number of elements in the input sequence. To address this problem, we use attention scheme \cite{Vinyals2015Pointer} model $p(C_l|C_1,\cdots,C_{l-1},E)$ as follows
\begin{equation}
\left\{ \begin{array}{l}
 u_l^t  = \beta ^T \tanh (W^1 s_l  + W^2 d_l ), \\
 p(C_l |C_1 , \ldots C_{l - 1} ,E) = {\rm{softmax}}(u^t ), \\
 \end{array} \right.
\end{equation}
where softmax normalizes $u^t$ to be an output distribution over the inputs. This approach targets problems whose outputs are correspond to positions in the input like the Seq2Seq learning problem for PDA design.

Traditional Seq2Seq models are often difficult to train due to the lack of an accurate assessment algorithm. In our problem, we can use Definition 1 as an evaluator to speed up the convergence of the algorithm. Reinforcement learning is used to achieve this. The environment state is the generated array $\overline{\textbf{P}}$, the action is the color assigned to each edge, the policy is the coloring scheme and the reward is defined as
\begin{equation}
R(\overline {\textbf{P}} ) = \left\{ \begin{array}{l}
 1,\;if\;\overline {\textbf{P}}\; is\;a\;PDA \\
  - 1,\;elsewise. \\
 \end{array} \right.
\end{equation}
Thus, the objective function of our model is
\begin{equation}
f = \frac{1}{{\rm K}}\sum\limits_{k = 1}^{\rm K} {R(\overline {\textbf{P}}_k )\log (C_k^E |E_k ;\theta )}.
\end{equation}
Using the $\rm K$ samples we update the parameters by using the gradient
\[
\nabla _\theta  f = \frac{1}{{\rm K}}\sum\limits_{k = 1}^{\rm K} {R(\overline {\textbf{P}}_k )\nabla _\theta  \log (C_k^E |E_k ;\theta )}
\]

We now present the attention-based Seq2Seq placement delivery network as Fig. 1. This network contains three main parts, placement phase, attention-based Seq2Seq model and delivery phase. In the first phase, the network output a adjacent matrix based on the parameters $K,F,Z$ and $S$. In the second phase, a proper coloring strategy is given and it the colors are used to construct a PDA. Finally, the delivery phase broadcasts data packets to the users using the PDA structured by the second phase. The detailed processes of this network are shown in Algorithm 1.

\begin{algorithm}
\caption{Seq2Seq Placement Delivery Network}
\label{alg2}
\begin{algorithmic}[1]
\STATE  \textbf{Training: } Training the attention-based Seq2Seq neural network $\mu$.\\
\STATE  \textbf{Placement: } Use the placement of AN scheme to broadcast some file packets and get matric $A$.\\
\STATE \textbf{Attention: } Use $\mu$ and $A$ to get the PDA $\overline{\textbf{P}}$.\\
\STATE \textbf{Delivery: } Use $\overline{\textbf{P}}$ and the delivery scheme of \cite{7973185} to broadcast the remaining packets.\\
\end{algorithmic}
\end{algorithm}

Note that the PDAs are scarce, only a small amount of training data can be obtained by using schemes proposed in \cite{7973185} and \cite{8080217}. Thus, we pre-train our deep neural network using the data generated by Theorem 1.

\section{SIMULATION RESULT AND DISCUSSION}
In this section, simulation results are provided to illustrate the effectiveness of the proposed method. Seq2Seq \cite{Sutskever2014Sequence} and Seq2Seq with attention \cite{8246560} are used as benchmarks for comparison.  We also compare the complexity with \cite{8080217}.

\begin{figure}
  \begin{minipage}[t]{0.45\linewidth}
    \centering
    \includegraphics[width=1.4in,height=1.8in]{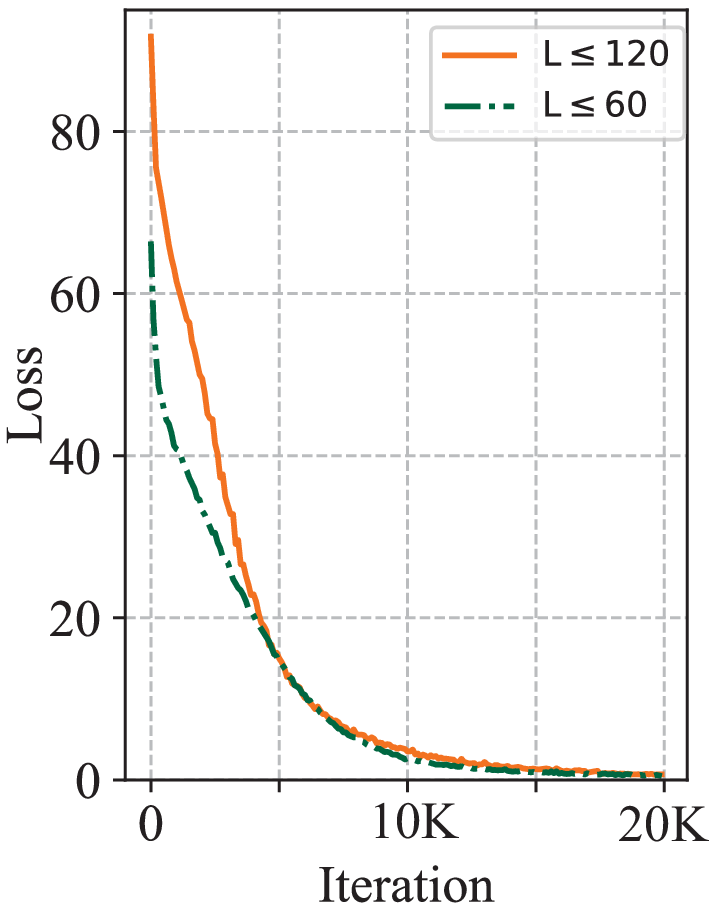}
    \caption{Training Loss}
  \end{minipage}%
  \begin{minipage}[t]{0.45\linewidth}
    \centering
    \includegraphics[width=1.4in,height=1.8in]{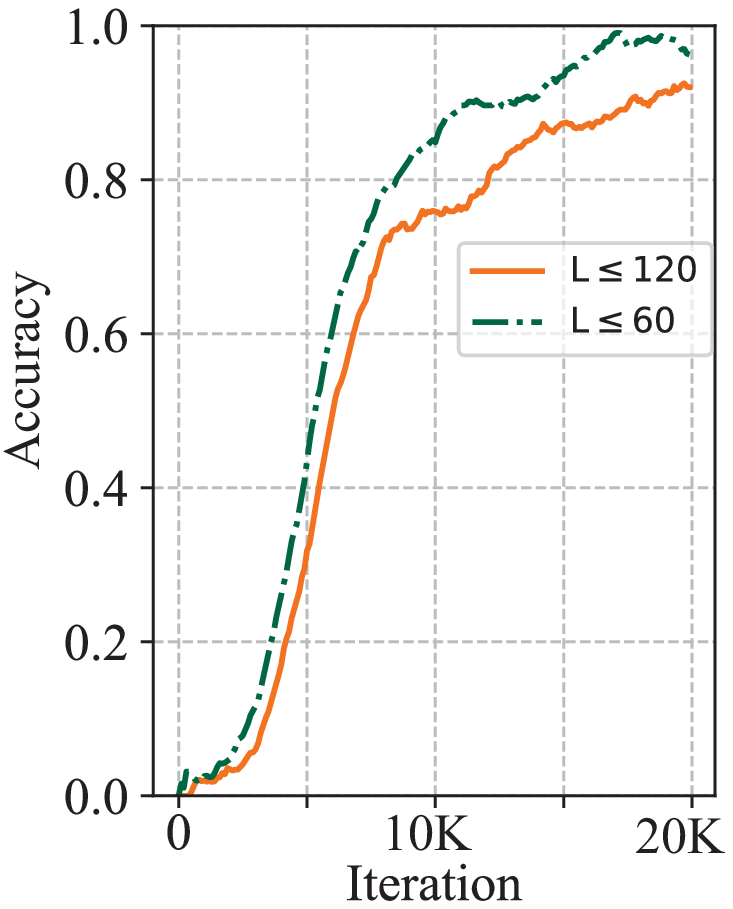}
    \caption{Training Accuracy}
  \end{minipage}%
\end{figure}

Fig. 2 shows the convergence behavior of the loss function. Fig. 3 shows the training accuracy of our model. We can find that no matter how $(K,F,M,N)$ is, as long as $L$ is less than 60 or 120, our algorithm shows good convergence performance and the training accuracy can reach more than 90\%.

Fig. 4 shows the result of solving the strong edge coloring problem of the $(21,7,3,4)$ caching system. Our solution use $33$ different color and it can be used to construct the optimal PDA.
\begin{figure}
\centering
\resizebox{7.6cm}{5.2cm}{\includegraphics{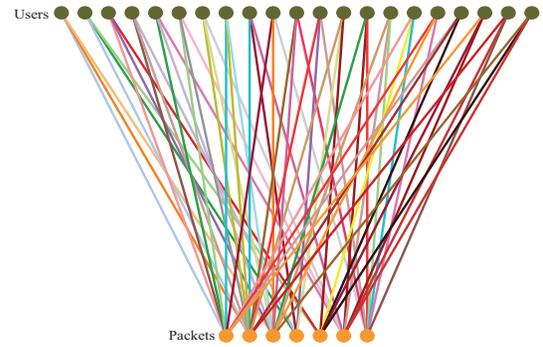}}
\caption{Strong Edge Coloring Result}
\end{figure}

\begin{table}
\captionsetup{format=plain, labelfont=bf,
  justification=raggedright, labelsep=newline}
\caption{NUMERICAL COMPARISONS}

  \centering
  \begin{tabular}{lp{1.8cm}p{1.8cm}p{1.8cm}}
  \hline
  Method &Test accuracy ($L\leq 60$)  &Test accuracy ($L\leq 120$) \\
  \hline
  Seq2Seq & 37.60\% & 28.43\%\\
  Seq2Seq with attention & 67.22\% & 65.59\% \\
  Our approach  & $\textbf{88.96}$\% & $\textbf{82.86}$\% \\
  \hline
  \end{tabular}
\end{table}

To use Seq2Seq or Seq2Seq with attention methods solve the learning problem for PDA design, the length of input and output should be fixed. Notice that this is not necessary for our method. The experimental results in Table 2 show that our method can achieve accuracy of $88.96\%$ and $82.86\%$ when $L$ is less than 60 and 120. It demonstrates that our proposed method is obviously better than the other two methods.

To construct a $(K,F,Z,S)$ PDA which corresponding to a colored bipartite graph $G(\mathscr{K},\mathscr{F},\mathscr{E})$, we compare the following 2 schemes:

 1) Scheme 1: Each users' group implements the strong coloring scheme in \cite{8080217}. The complexity of this method is ${\cal{O}} \left( {(K + F)\Delta \left| \mathscr{E} \right|^2 } \right)$, where $\Delta \leq 2$ is the degree of the graph $G$ and $\left| \mathscr{E} \right|$ is the size of the edge set.

 2) Scheme 2: Each user's group implements our method. If the attention-based Seq2Seq neural network has been trained, the complexity of our PDA design method is ${\cal{O}}\left( {\left| \mathscr{E} \right|\log (\left| \mathscr{E} \right|)} \right)$.

We can find that our method has a low complexity and can achieve approximate solutions to PDA design.

\section{Conclusion}
In this paper, we established the connection between the placement delivery array in coded caching and the sequence-to-sequence learning. We first proposed a learning method to construct PDAs using attention model and reinforcement learning. Then a new coded caching scheme is constructed based on the deep neural architecture. Numerical results demonstrated that the proposed method can effectively implement coded caching and the complexity is low. Future work is going to consider a more efficient deep neural network by using Generative Adversarial Networks (GAN) \cite{Goodfellow2014Generative} and achieve distributed system by using coded computing \cite{CodedComputing}.

% or


%\appendices
%% you can choose not to have a title for an appendix
%% if you want by leaving the argument blank
%\section{}
%% use section* for acknowledgment
%\section*{Acknowledgment}

\ifCLASSOPTIONcaptionsoff
  \newpage
\fi



%\bibliographystyle{IEEEtran}

\bibliographystyle{IEEEtran}
\bibliography{IEEEabrv,PAD}
%\begin{thebibliography}{1}
%\end{thebibliography}


%\end{IEEEbiography}
%% insert where needed to balance the two columns on the last page with
%%\newpage


%\vfill


\end{document}